\documentclass[prb,twocolumn,floatfix,showpacs,preprintnumbers,superscriptaddress,amsmath,amssymb]{revtex4}

\usepackage{graphicx}
\usepackage{dcolumn}
\usepackage{bm}

\begin{document}

\title{Theoretical study of the conductance of ferromagnetic atomic-sized contacts}

\author{M. H\"afner}
\affiliation{Institut f\"ur Theoretische Festk\"orperphysik,
Universit\"at Karlsruhe, D-76128 Karlsruhe, Germany}

\author{J.K. Viljas}
\affiliation{Institut f\"ur Theoretische Festk\"orperphysik,
Universit\"at Karlsruhe, D-76128 Karlsruhe, Germany}
\affiliation{Forschungszentrum Karlsruhe, Institut f\"ur Nanotechnologie, D-76021
Karlsruhe, Germany}

\author{D. Frustaglia}
\affiliation{NEST-CNR-INFM \& Scuola Normale Superiore, I-56126 Pisa, Italy}

\author{F. Pauly}
\affiliation{Institut f\"ur Theoretische Festk\"orperphysik,
Universit\"at Karlsruhe, D-76128 Karlsruhe, Germany}
\affiliation{Forschungszentrum Karlsruhe, Institut f\"ur Nanotechnologie, D-76021
Karlsruhe, Germany}

\author{M. Dreher}
\affiliation{Fachbereich Physik, Universit\"{a}t Konstanz, D-78457 Konstanz,
Germany}

\author{P. Nielaba}
\affiliation{Fachbereich Physik, Universit\"{a}t Konstanz, D-78457 Konstanz,
Germany}

\author{J.C. Cuevas}
\affiliation{Departamento de F\'{\i}sica Te\'orica de la Materia
Condensada, Universidad Aut\'onoma de Madrid, E-28049 Madrid, Spain}
\affiliation{Institut f\"ur Theoretische Festk\"orperphysik,
Universit\"at Karlsruhe, D-76128 Karlsruhe, Germany}
\affiliation{Forschungszentrum Karlsruhe, Institut f\"ur Nanotechnologie, D-76021
Karlsruhe, Germany}

\date{\today}

\begin{abstract}
Recently, different experiments on the transport through atomic-sized contacts
made of ferromagnetic materials have produced contradictory results. In particular,
several groups have reported the observation of half-integer conductance quantization,
which requires having full spin polarization and perfectly conducting channels.
Motivated by these surprising results, we have studied theoretically the conductance
of ideal atomic contact geometries of the ferromagnetic 3$d$ materials Fe, Co, and Ni 
using a realistic tight-binding model. Our analysis shows that in the absence of magnetic 
domains, the $d$ bands of these transition metals play a key role in the electrical 
conduction. In the contact regime this fact has the following important consequences 
for the three materials: (i) there are partially open conduction channels and therefore 
conductance quantization is not expected, (ii) the conductance of the last plateau is 
typically above $G_0=2e^2/h$, (iii) both spin species contribute to the transport and thus 
there is in general no full current polarization, and (iv) both the value of the conductance 
and the current polarization are very sensitive to the contact geometry and to disorder. 
In the tunneling regime we find that a strong current polarization can be achieved.
\end{abstract}


\maketitle

\section{Introduction\label{introduction}}
Metallic nanowires fabricated by means of scanning-tunneling microscope
and break-junction techniques have turned out to be a unique playground 
to test basic concepts of electronic transport at the atomic scale.\cite{Agrait2003}
Usually the conductance of these contacts is described by the Landauer formula
$G = G_0 \sum_n T_n$, where the sum runs over all the available conduction 
channels, $T_n$ is the transmission for the $n$th channel, and $G_0 =2e^2/h$ 
is the quantum of conductance. In the case of broken spin symmetry
one should include a sum over spin and replace $G_0$ by $e^2/h$ in the
previous formula. It has been shown that the number of channels in a one-atom 
contact is mainly determined by the number of valence orbitals of the central atom, 
and the transmission of each channel is fixed by the local atomic
environment.\cite{Cuevas1998a,Scheer1998,Cuevas1998b} Thus, for instance, a 
one-atom contact of a monovalent metal such as Au sustains a single channel, while 
for $sp$-like metals such as Al or Pb one finds three channels due to the 
contribution of the $p$ orbitals. More importantly for the discussion in this
work, in a transition metal such as Nb the contribution of the $d$ orbitals leads
to five partially open channels.\cite{Cuevas1998a,Scheer1998,Ludoph2000a}

In the last years a lot of attention has been devoted to the experimental analysis of
contacts of magnetic materials.\cite{Sirvent1996,Costa1997,Hansen1997,
Ott1998,Oshima1998,Komori1999,Ono1999,Garcia1999,Ludoph2000b,Viret2002,Elhoussine2002,
Shimizu2002,Gillingham2002,Rodrigues2003,Gillingham2003a,Gillingham2003b,Untiedt2004,Gabureac2004,
Yang2004,Costa2005,Bolotin2006a,Keane2006,Bolotin2006b,Viret2006} In particular, several 
groups have reported the observation of half-integer conductance quantization
in the last stages of the breaking of nanowires.\cite{Elhoussine2002,Shimizu2002,
Gillingham2002,Rodrigues2003,Gillingham2003a,Gillingham2003b}
Moreover, some authors have shown that the conductance histograms are very sensitive 
to a magnetic field. For instance, Ono \emph{et al.}~\cite{Ono1999} reported 
that the peaks at $1G_0$ and $2G_0$ which are observed without magnetic field are joined by 
additional peaks near $0.5G_0$ and $1.5G_0$ for fields above 5 mT.
These findings are rather surprising and challenge our present understanding
of the transport properties of atomic-sized contacts. Half-integer conductance quantization
would require to have fully spin-polarized contacts and conduction channels
with perfect transparency. Neither of these properties are expected in the 3$d$ ferromagnetic
materials (Fe, Co, and Ni). In these transition metals both the majority spin and 
the minority spin electrons at the Fermi energy contribute to transport and, moreover,
the partially occupied $d$ orbitals are expected to give rise to partially open
channels. Finally, it is impossible in practice for a magnetic field to lift the spin degeneracy
in a metallic contact. A field of 10 T produces a Zeeman splitting on the order of 1 meV,
which is extremely small in comparison with the bandwidth of even a single-atom contact,
this being at least on the order of several electronvolts.\cite{Cuevas1998a,Cuevas1998b}
In this sense, in the absence of magnetic domains and magnetostriction, it is difficult
to understand how an external magnetic field could significantly modify the conductance of an
atomic contact.

More recently, Untiedt \emph{et al.}~\cite{Untiedt2004} measured the conductance for contacts
of several magnetic metals (Fe, Co, and Ni) using break junctions at low temperatures
and under cryogenic vacuum conditions. They reported the absence of fractional
conductance quantization, even when a high magnetic field was applied, which agrees better with 
the picture described above. Several recent model calculations support these 
findings.\cite{Martin2001,Smogunov2002,Delin2003,Velev2004,Bagrets2004,Rocha2004,
Jacob2005,Wierzbowska2005,Dalgleish2005,Jacob2006} These studies have convincingly 
shown that neither conductance quantization nor full spin polarization are to be expected in 
contacts of this kind. 

In spite of the coherent picture that is emerging, one still misses in the 
literature a comparative analysis of the ferromagnetic 3$d$ materials (Fe, Co, and Ni)
that clarifies basic issues like which orbitals are relevant for the transport, the role
of atomic disorder, or the dependence of the spin polarization of the current on 
the thickness of the contact. To fill this gap, we present in this work detailed 
calculations of the conductance of ferromagnetic atomic contacts of Fe, Co, and Ni. 
In our calculations we study ideal contact geometries using a realistic tight-binding 
model. We only consider situations where the contact region is a single magnetic domain. 
Our results indicate that for a few-atom contact of the three materials one can draw the 
following general conclusions: (i) there is no conductance quantization, mainly due to 
the partially open conduction channels of the minority spin electrons, (ii) the last 
plateau has typically a conductance above $G_0=2e^2/h$, (iii) the current is not
fully spin-polarized and both spin species contribute to the transport, and (iv) both
the value of the conductance and the current polarization are very sensitive to the 
contact geometry and disorder. The origin of all these findings can be traced back to
the fact that the $d$ bands of these transition metals play a very important role 
in the electrical conduction. This is in contrast with other physical situations
such as tunnel junctions or bulk systems. Finally, we find that in the tunnel regime,
which is reached when the contacts are broken, the nature of the conduction changes 
qualitatively and almost fully spin-polarized currents are indeed possible.

The rest of the paper is organized as follows. In the following section we describe
our tight-binding approach to compute the conductance of the ferromagnetic atomic
contacts. Section \ref{ideal_geometries} is devoted to the analysis of the results of the
conductance of representative one-atom thick contacts of Fe, Co, and Ni. Moreover,
we include in this section a discussion of the conductance in the tunnel regime. 
In Sec.~\ref{disorder_geometries} we discuss the influence of atomic disorder on the conductance of 
single-atom contacts. Finally, in Sec.~\ref{conclusions} we summarize and discuss the main results.

\begin{figure*}[t]
\includegraphics*[width=0.90\textwidth,clip]{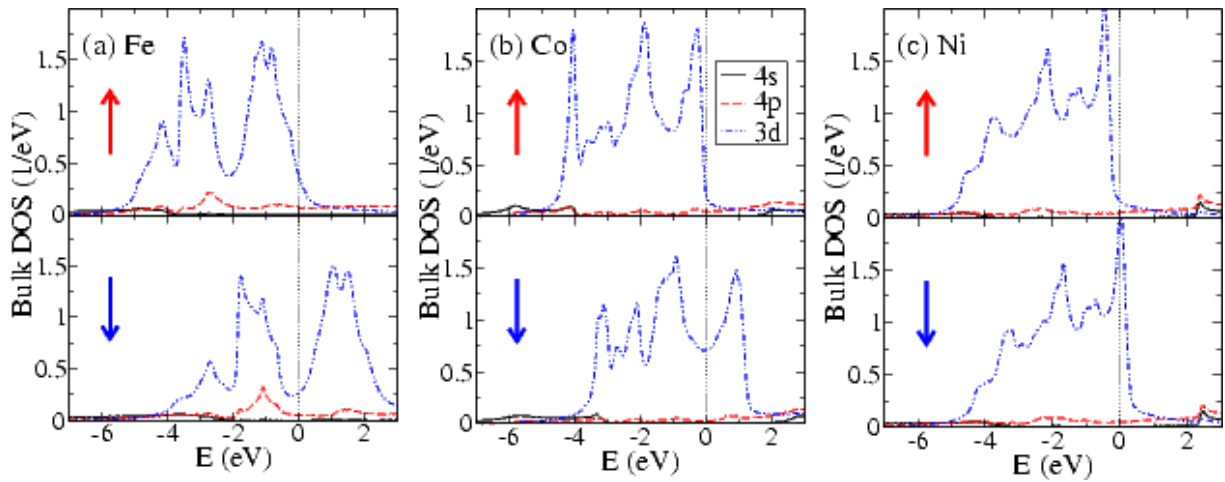}
\caption{\label{bulk-dos} (Color online) Bulk density of states (DOS) of Fe, Co, and Ni, resolved
with respect to the individual contributions of 3$d$, 4$s$, and 4$p$ orbitals, as indicated in the 
legend. Furthermore, the upper panels show the DOS for the majority spins and the lower ones the DOS 
for minority spins. The Fermi energy is set to zero and it is indicated by the vertical dashed line.}
\end{figure*}

\section{Description of the theoretical model\label{model}}

Our goal is to compute the low-temperature linear conductance of atomic-sized contacts 
of the 3$d$ ferromagnetic metals Fe, Co, and Ni. For this purpose, we use a tight-binding 
model based on the sophisticated parametrization introduced in Ref.~[\onlinecite{Mehl1996}].
Such tight-binding models have been successful in the description of electron transport 
in metallic atomic contacts.\cite{Cuevas1998a,Cuevas1998b,Haefner2004} Our approach 
follows closely the one used in Refs.~[\onlinecite{Brandbyge1999,Dreher2005,Viljas2005}], 
and we now proceed to describe it briefly.

In our approach the electronic structure of the atomic contacts is described in terms of 
the following tight-binding Hamiltonian written in a non-orthogonal basis
\begin{equation}
\hat{H} = 
\sum_{i \alpha, j \beta, \sigma} H^{\sigma}_{i\alpha,j\beta}
\hat c^{\dagger}_{i \alpha,\sigma} \hat c_{j \beta,\sigma} .
\label{Hamiltonian}
\end{equation}
\noindent
Here $i,j$ run over the atomic sites, $\alpha,\beta$ denote the different atomic 
orbitals, and $\sigma=\uparrow,\downarrow$ denotes the spin. 
Furthermore, $H^{\sigma}_{i \alpha,j \beta}$ for $i=j$ and $\alpha=\beta$ are the spin-dependent 
on-site energies, and for $i\neq j$ the hopping elements, while 
$H^{\sigma}_{i \alpha, i \beta}=0$ for $\alpha\neq\beta$. 
In addition, we need the overlaps between the different orbitals, $S_{i\alpha,j\beta}$, 
which are spin-independent. We take all these parameters from the bulk parametrization 
of Ref.~[\onlinecite{Mehl1996}], which is known to accurately reproduce the band 
structure and total energy of bulk ferromagnetic materials.\cite{Bacalis2001} Notice that
in our model there is no mixing of the two spin species, which means that, in particular, 
we do not consider spin-orbit interaction. The atomic basis is formed by 9 orbitals 
($3d,4s,4p$), which give rise to the main bands around the Fermi energy in Fe, Co, and Ni. 
It is important to emphasize that in this parametrization both the hopping elements and 
the overlaps are functions of the relative positions of the atoms, which allows us to study 
also geometrical disorder. These functions have a cut-off radius that encloses atoms well 
beyond the 10th nearest neighbors in a bulk geometry for Fe, Co, and Ni.

In order to compute the linear conductance, we apply a standard Green functions method 
[\onlinecite{Cuevas1998a,Haefner2004,Brandbyge1999,Dreher2005,Viljas2005}]. For this we 
divide the system into three parts, the left ($L$) and right ($R$) leads, and the central cluster 
($C$) containing the constriction. In this way, the retarded central cluster Green functions, 
${\bf G}^{\sigma}_{CC}$, are given by
\begin{equation}
{\bf G}^{\sigma}_{CC}(E) = \left[ E {\bf S}_{CC} - {\bf H}^{\sigma}_{CC} -
{\bf \Sigma}^{\sigma}_{L} (E) - {\bf \Sigma}^{\sigma}_{R} (E) \right]^{-1} ,
\end{equation}
\noindent
Here ${\bf H}^{\sigma}_{CC}$ and ${\bf S}_{CC}$ are the Hamiltonian and the overlap 
matrix of the central cluster, respectively, and
${\bf \Sigma}^{\sigma}_{L/R}$ are the self-energies, which contain the information 
of the electronic structure of the leads and their coupling to the central part of 
the contact. 
These self-energies can be expressed as
\begin{equation}
{\bf \Sigma}^{\sigma}_{L}(E) = \left( {\bf H}^{\sigma}_{CL} - 
E {\bf S}_{CL} \right) {\bf g}^{\sigma}_{LL}(E) 
\left( {\bf H}^{\sigma}_{LC} - E {\bf S}_{LC} \right) ,
\end{equation}
\noindent
with a similar equation for ${\bf \Sigma}^{\sigma}_{R}(E)$.
Here, for example,  ${\bf H}^{\sigma}_{CL}$ is the hopping matrix
connecting the central cluster $C$ and the lead $L$, ${\bf S}_{CL}$ is the 
corresponding block of the overlap matrix, and ${\bf g}^{\sigma}_{LL}(E)$ 
is the retarded Green function of the uncoupled lead. Both infinite leads are described by 
ideal surfaces, the Green functions of which are calculated within the same 
tight-binding parametrization using the decimation technique described in 
Ref.~[\onlinecite{Guinea1983}]. 

In an atomic contact the local environment in the region of the constriction is very
different from that of the bulk material. In particular, this fact can lead to
large deviations from the approximate local charge neutrality that typical metallic
elements exhibit. We correct this problem by imposing charge neutrality on all
the atoms of the nanowire through a self-consistent variation of ${\bf H}^{\sigma}_{CC}$,
following Ref.~[\onlinecite{Brandbyge1999}] and shifting both spin species equally.

The linear conductance at low temperature can now be expressed in terms of 
the Landauer formula
\begin{equation}
G = \frac{e^2}{h} \sum_{\sigma} T_{\sigma}(E_F) ,
\end{equation}
where $T_\sigma(E)$ is the total transmission for spin 
$\sigma=\uparrow,\downarrow$ at energy $E$, and $E_F$ is the Fermi energy.
We also define the spin-resolved conductances $G_\sigma=(e^2/h)T_\sigma(E_F)$,
such that $G=G_{\uparrow}+G_{\downarrow}$.
The transmissions are obtained as follows
\begin{equation}
T_{\sigma}(E) = \mbox{Tr} [ {\bf t}_{\sigma}(E) 
{\bf t}_{\sigma}^{\dagger}(E) ] = \sum_{n} T_{n,\sigma}(E) ,
\end{equation}
where ${\bf t}_{\sigma}(E)$ is the transmission matrix and 
$T_{n,\sigma}(E)$ are the individual transmission eigenvalues 
for each spin $\sigma$.
The transmission matrix can be calculated in 
terms of the Green functions ${\bf G}^{\sigma}_{CC}$ as follows
\begin{equation}
{\bf t}_{\sigma}(E) = 2 \; \left[{\bf \Gamma}^{\sigma}_{L}(E)\right]^{1/2} {\bf G}^{\sigma}_{CC}(E)
\left[ {\bf \Gamma}^{\sigma}_{R} (E) \right]^{1/2} .
\end{equation}
\noindent
Here, ${\bf \Gamma}^{\sigma}_{L/R}(E)$ are the scattering rate matrices given by
${\bf \Gamma}^{\sigma}_{L/R}(E) = -\mbox{Im} [ {\bf \Sigma}^{\sigma}_{L/R} (E) ]$.

\begin{figure*}[t]
\includegraphics*[width=0.92\textwidth,clip]{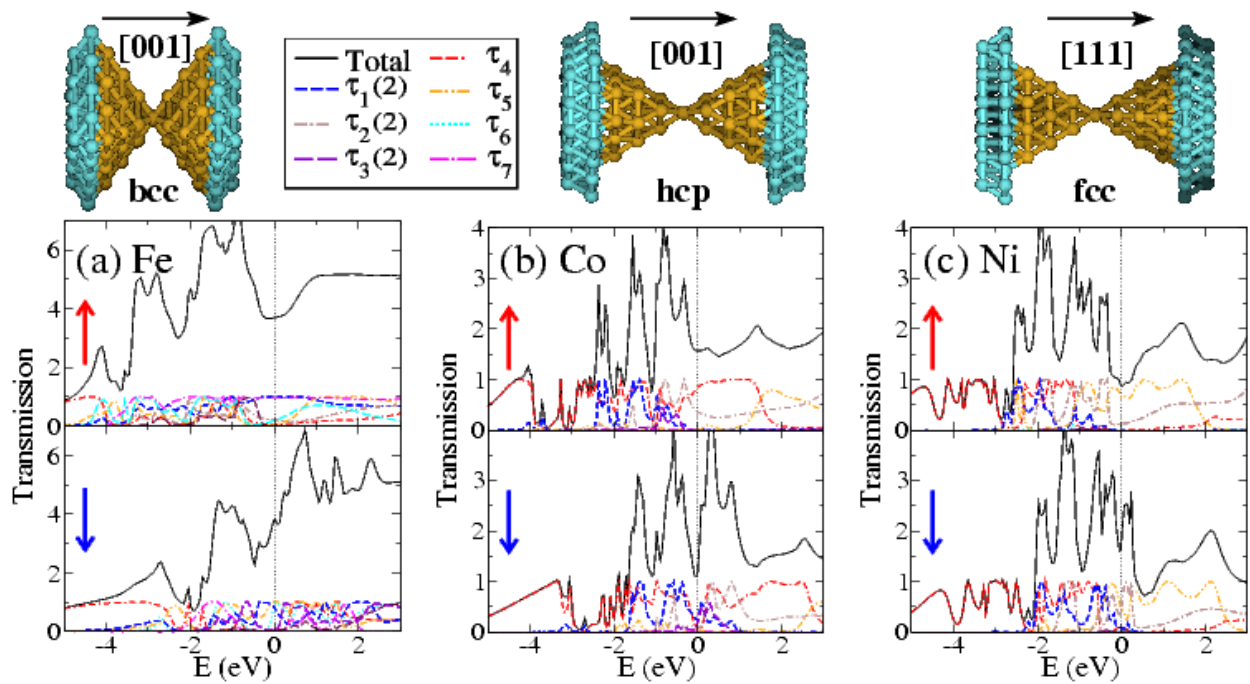}
\caption{\label{single-atom} (Color online) Transmission as a function of energy for 
the three single-atom contacts of (a) Fe, (b) Co, and (c) Ni, which are shown in the 
upper panels. We present the total transmission (black solid line) for both majority spins
and minority spins as well as the transmission of individual conduction channels 
that give the most important contribution at Fermi energy, which is indicated by a vertical 
dotted line. The blue, brown and violet dash-dotted lines of $\tau_1$, $\tau_2$, and $\tau_3$ 
refer to twofold degenerate conduction channels. The legends in the upper graphs 
indicate, in which direction the contacts are grown. These contacts contain in the central 
region 59 atoms for Fe, 45 for Co, and 39 for Ni. The blue atoms represent a part of the 
atoms of the leads (semi-infinite surfaces) that are coupled to the central atoms in our model.} 
\end{figure*}

\section{Conductance of ideal single-atom contacts of Fe, Co, and Ni\label{ideal_geometries}}

The goal of this section is the analysis of the conductance of ideal, yet plausible
one-atom contact geometries for the three ferromagnetic metals (Fe, Co, and Ni)
considered in this work. In order to understand the results described below,
it is instructive to first discuss the bulk density of states (DOS). 
The spin- and orbital-resolved bulk DOS of these materials around $E_F$, 
as calculated from our tight-binding
model, is shown in Fig.~\ref{bulk-dos}. The common feature for
the three ferromagnets is that the Fermi energy for the minority spins lies inside 
the $d$ bands. This fact immediately suggests \cite{Cuevas1998a,Scheer1998} 
that the $d$ orbitals may play an important role in the transport. Occupation
of the $s$ and $p$ orbitals for both spins is around 0.25 and 0.4 electrons, respectively. 
For the majority spins the Fermi energy lies close to the edge of the $d$ band. The 
main difference between the materials is that for Fe there is still an important 
contribution of the $d$ orbitals, while for Ni the Fermi level is in a region where 
the $s$ and $p$ bands become more important. The calculated values of the magnetic moment per atom 
(in units of the Bohr magneton) of $2.15$ for Fe, $1.3$ for Co, and $0.45$ for Ni are reasonably 
close to literature values.\cite{CRC1999}

We now proceed to analyze in detail the conductance of some ideal one-atom geometries,
which are chosen to simulate what happens in the last conductance plateau before the
breaking of the nanowires. First, we consider the one-atom contacts shown in the upper panels
of Fig.~\ref{single-atom}. These geometries are constructed starting with a single atom 
and choosing the nearest neighbors in the next layers of the ideal lattice along the 
direction indicated with an arrow. In the case of Fig.~\ref{single-atom}, for Fe (bcc lattice 
with a lattice constant of $2.86$ \AA) the contact is grown along the [001] direction, for 
Co (hcp lattice, lattice constant $2.51$ \AA) along the [001] direction (parallel to  ``$c$ axis") 
and for Ni (fcc lattice and lattice constant $3.52$ \AA) along the [111] direction. The number 
of atoms in the central region has been chosen large enough, such that the transmission does 
not depend anymore on the number of layers included. Moreover, as explained in the previous section,
the central region is coupled seamlessly to ideal surfaces grown along the same direction. 

Let us start describing the results for the Fe one-atom contact of Fig.~\ref{single-atom}(a).
There we present the total transmission for majority spins and minority spins as a function 
of energy as well as the individual transmissions.
We find for this particular geometry the spin-resolved conductances 
$G_{\uparrow} = 3.70e^2/h$  ($\uparrow$ for majority spins) and $G_{\downarrow} = 3.75e^2/h$ 
($\downarrow$ for minority spins), which results in a total 
conductance of $3.7G_0$. The conductance $G_\uparrow$ for the majority spins 
is the result of up to 8 open channels (with a transmission higher than 0.01), while 
for the minority spins there are 11 channels giving a significant contribution to $G_\downarrow$. 
The large number of channels
and consequently the high conductance, are partially due to the large apex angle of $71^\circ$ of the
pyramids. As a consequence of this, the layers next to the central atom couple to each other
and give rise to a significant tunnel current that proceeds directly without traversing the
central atom. On the other hand, the larger number of channels for the minority spins is due to the 
key contribution of the $d$ orbitals that dominate the transport through this spin species, while
for the majority spins the $s$ and $p$ orbitals are the more relevant ones. This fact, which is supported
by the analysis of the local density of states (not shown here), is a simple consequence of
the position of the Fermi energy and the magnitude of the spin splitting (see discussion of
the bulk DOS above).

We define the spin polarization $P$ of the current as 
\begin{equation}
P = \frac{G_{\uparrow}-G_{\downarrow}}{G_{\uparrow}+G_{\downarrow}} .
\end{equation}
With this definition we find a value of $P = -0.7\%$ for the Fe 
one-atom contact of Fig.~\ref{single-atom}(a). In order to compare to the polarization 
of bulk we have calculated the transmission at the Fermi energy for a series of contact 
geometries where a bar of constant diameter bridges the two lead surfaces.  
When the diameter of the bar (or central region of the contact) is increased, 
the polarization grows continously and saturates at a value of $P = +40\%$ for a contact 
containing $219$ atoms in 7 layers. This is in good agreement with the 
experimental value obtained using normal-metal-superconductor point contacts.\cite{Soulen1998}
Notice that $P$ can be quite different in an atomic contact as compared to
bulk. This is because the conductance is not simply controlled by the DOS
at the Fermi energy, but the precise coupling between the orbitals in the constriction 
plays a crucial role.

\begin{figure*}[t]
\includegraphics*[width=0.92\textwidth,clip]{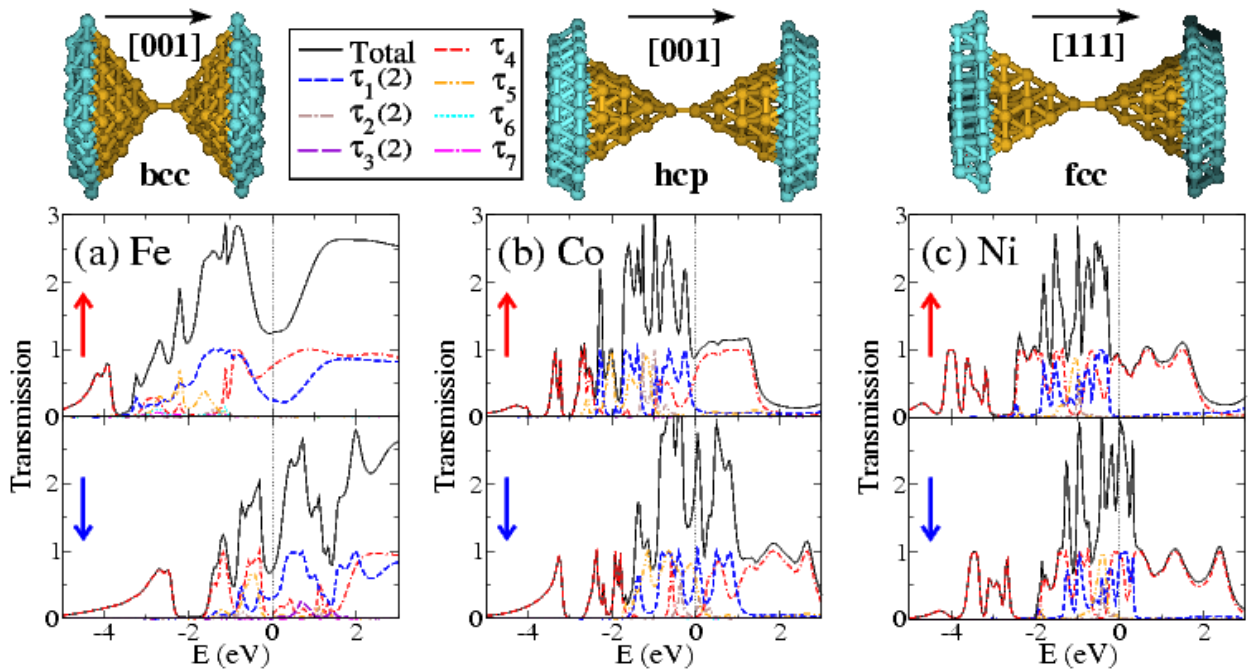}
\caption{\label{dimer} (Color online) The same as in Fig.~\ref{single-atom} but for the geometries
shown in the upper graphs, which contain a dimer in the central part of the 
contact. The two dimer atoms are at the bulk nearest neighbor distance from each other.}
\end{figure*}

For the Co contact depicted in Fig.~\ref{single-atom}(b) the transmission is lower 
than for Fe, partly due to the smaller apex angle of the hcp pyramids. In this case we find
$G_{\uparrow} = 1.57e^2/h$ for majority spins and $G_{\downarrow} = 1.21e^2/h$ for minority spins, 
summing up to a total conductance 
of $1.4G_0$. There are 3 channels contributing to $G_{\uparrow}$ 
and 8 channels to $G_{\downarrow}$. As in the case of Fe, the larger
number of channels for the minority spins is due to the position of the Fermi level and the
resulting contribution of the $d$ orbitals for this spin. We also find that there is
a small but non-negligible contribution of channels that proceed directly without
crossing the central atom. This explains, in particular, why one has 8 channels
for the minority spins, although at most 6 bands ($s$ and $d$) have a significant DOS at this
energy. Turning to the current polarization, we find a value of $P = +13\%$ for the Co 
one-atom contact. We also calculate the polarization for a series of Co bars 
with increasing diameter in hcp $[001]$ direction.  As the diameter increases, 
the polarization decreases to a value 
of $P = -41\%$ for a contact containing five layers of $37$ atoms each, again in good 
agreement with the experiment.\cite{Soulen1998} Notice again that not only the magnitude 
of $P$ for a one-atom contact can be quite different from bulk, but also its sign can
be the opposite.

Finally, the Ni contact shown in Fig.~\ref{single-atom}(c) exhibits conductances of
$G_{\uparrow} = 0.85e^2/h$ for majority and $G_{\downarrow} = 1.80e^2/h$ for minority spins, 
adding up to a total conductance of $1.3G_0$. The $G_\uparrow$ consists of 3 channels, 
due to the contribution of the $s$ and $p$ orbitals, and $G_{\downarrow}$ contains 6 channels,
which originate from the contribution of the $d$ orbitals. In this case we find a 
value for the polarization of $P = -34\%$. Once more we have investigated the polarization of bulk Ni 
in a series of large Ni bars in fcc $[111]$ direction. Interestingly, the polarization decreases 
from $P = +3\%$ for a contact of $28$ atoms in four layers to $P = -41\%$ for a contact 
consisting of $244$ atoms in four layers.

Now we turn to the analysis of the geometries shown in the upper panels of 
Fig.~\ref{dimer}. The difference with respect to the geometries of Fig.~\ref{single-atom}
is the presence of a dimer in the central part of the contacts. This type of
geometry has frequently been observed in molecular dynamics simulations of atomic 
contacts of Al (Ref.~[\onlinecite{Jelinek2003}]) and Au (Ref.~[\onlinecite{Dreher2005}]) 
and we also find them in our simulations of Ni contacts in the last stages of the breaking 
process.\cite{Pauly2006}

\begin{figure*}[t]
\includegraphics*[width=0.98\textwidth,clip]{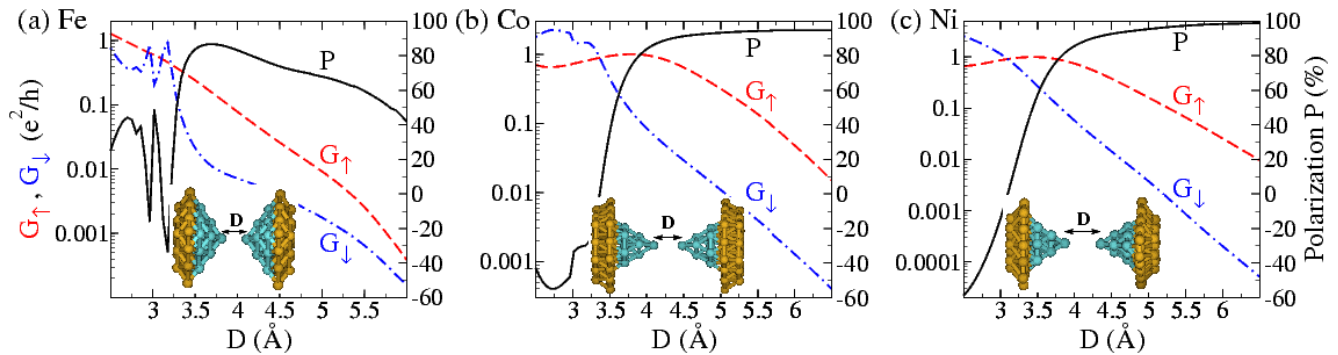}
\caption{\label{tunnel} (Color online) Conductance over tip separation $D$ of similar geometries 
as in Fig.~\ref{dimer}. The conductance of majority spin ($G_{\uparrow}$, dashed lines and left scales) 
and minority spin ($G_{\downarrow}$, dash-dotted lines and left scales) is shown, together with the 
resulting current polarization (solid lines and right scales).}
\end{figure*}

Inserting a dimer in the geometries of Fig.~\ref{single-atom} results in a larger separation 
of the pyramids to the left and right of the central atom and therefore in a weaker
coupling between the layers next to the dimer. This is particularly important in the case of Fe.
The resulting transmission for the Fe contact with a central dimer is shown in Fig.~\ref{dimer}(a), 
where one can see that only 3 channels remain for the majority spins, yielding $G_{\uparrow} = 1.24e^2/h$,
while for the minority spins 3 channels contribute to $G_{\downarrow} = 0.70e^2/h$. The total 
conductance is $1.0G_0$ and 
the polarization $P = +28\%$. For Co the contact of Fig.~\ref{dimer}(b) with a central dimer 
exhibits $G_{\uparrow} = 0.90e^2/h$ and $G_{\downarrow} = 2.23e^2/h$, 
summing up to a total conductance of $1.6G_0$. 
The transmission is formed by 3 channels for the majority spins (with one clearly dominant) 
and 6 channels for minority spins and polarization is $P = -42\%$.
Finally, for the Ni contact in Fig.~\ref{dimer}(c) with a central dimer, a single channel 
contributes to $G_{\uparrow} = 0.86e^2/h$ and 4 channels add
up to $G_{\downarrow} = 2.66e^2/h$. This means that one has a total conductance of $1.8G_0$, 
while the current polarization adopts a value of $P = -51\%$. 

Beyond the precise numerical values detailed in the previous paragraph, we would like
to stress the following conclusions from the analysis of Fig.~\ref{dimer}. First, the transport
contribution of the minority spins is dominated by the $d$ orbitals, which
give rise to several channels (from 3 to 5 depending on the material). Second, for
the majority spins there is a smaller number of channels ranging from 3 for Fe to 1
for Ni. This contribution is dominated by the $d$ and $s$ orbitals for Fe and only by
the $s$ orbitals for Co and Ni. The relative contribution and number of channels of the
two spin species is a simple consequence of the position of the Fermi level and the 
magnitude of the spin splitting. In particular, notice that as we move from Fe to 
Ni, the Fermi energy lies more and more outside of the $d$ band for the majority spins, which
implies that the number of channels is reduced for this spin species. In particular,
for Ni a single majority spin channel dominates. On the other hand, notice that the
conductance values for the different contacts lie typically above $G_0$, which is
precisely what is observed experimentally.\cite{Untiedt2004}

So far, we have analyzed geometries for the so-called contact regime where the nanowires
are formed. As shown above, in this case the contribution of the $d$ bands makes it 
difficult to obtain large values of the current polarization. In this sense, one may wonder 
what happens in the tunnel regime when the contact is broken. 
In order to address this issue, we have simulated the breaking 
of the contacts by progressively separating the electrodes of the dimer geometries of
Fig.~\ref{dimer}. In this way, we have computed the conductance and the current polarization 
as a function of the tip separation $D$ and the results for the three materials are summarized 
in Fig.~\ref{tunnel}. With increasing $D$ one enters the tunnel regime, which is characterized
by an exponential decay of the conductance. In the regime shown in the graphs, Fe does not yet 
exhibit an exponential decay. In contrast, the conductances for Co and Ni are well fitted 
by an exponential $\exp(-\beta D)$ with $\beta=2.3$ \AA$^{-1}$ and $\beta=1.9$ \AA$^{-1}$, 
respectively. These values are in reasonable agreement with the WKB approximation\cite{Messiah}, 
which yields $\beta=2.2$ \AA$^{-1}$
using a work function of $5$ eV.\cite{CRC1999} Notice that deep 
in the tunnel regime for the three materials, the conductance for the majority spins largely 
overcomes the value of the minority spin conductance. This results in positive values of the 
current polarization $P$ and, in particular, for Co and Ni it reaches values very close to 100\%. 

\begin{figure*}[t]
\includegraphics*[width=0.94\textwidth,clip]{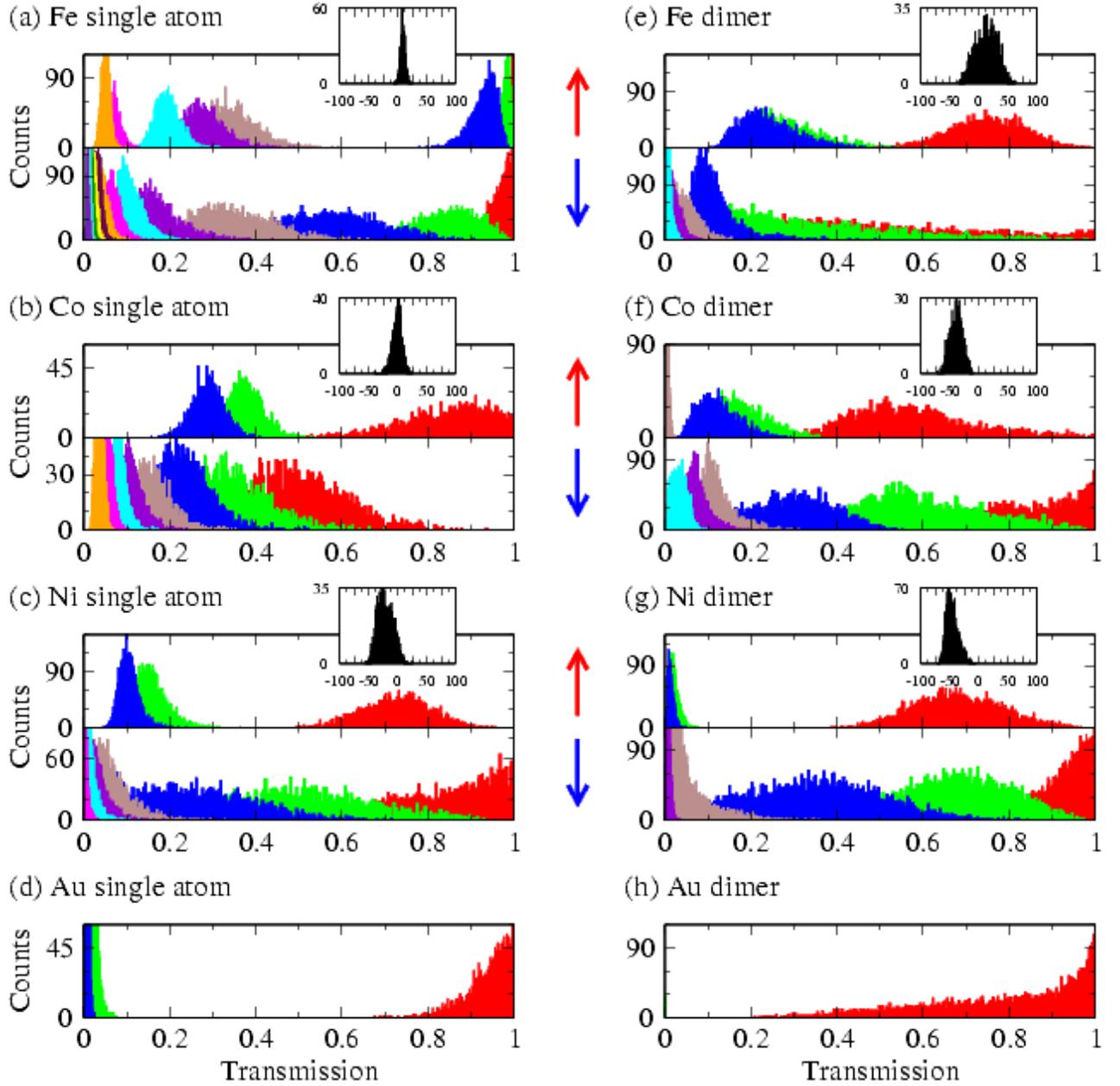}
\caption{\label{stat} (Color online) Histograms of transmission channels at Fermi energy, 
$T_{n,\sigma}(E_F)$, for 3000 perturbed realizations of ideal contact geometries of Fe, Co, and Ni. 
Panels (a)-(d) show histograms for contacts with a single central atom as in 
Fig.~\ref{single-atom} and panels (e)-(h) for contacts with a central dimer as in 
Fig.~\ref{dimer}. In panels (a)-(c) and (e)-(g) results for ferromagnetic 
contacts are presented: the upper parts of the panels refer to 
majority spin channels and the lower parts to minority spin channels. Only channels 
that contribute more than $0.01$ to transmission are displayed, and the histograms for smaller 
transmission values are in the front. The insets for the ferromagnetic materials show 
corresponding histograms for the current polarization $P$, where on the $x$ axis $P$ is 
given in $\%$. Finally, panels (d) and (h) show comparison histograms for 
fcc-Au calculated with a similar set of geometries as for Ni.}
\end{figure*}

The origin of these huge values of current polarization in the tunnel regime is the
following. In this regime the current is, roughly speaking, a convolution of the local 
densities of states on the tips weighted with the squared hoppings of the relevant orbitals of both 
electrodes. The hoppings between $3d$ orbitals decay faster with the separation of 
the tips than the corresponding hoppings of the $4s$ orbitals. Due to this faster 
decay, the conduction is then dominated by the $s$ orbitals and since the on-site energy 
for the minority spins lies further away from the Fermi energy than the corresponding one for majority spin,
the transmission through the latter one is much higher giving rise to a very large positive
spin polarization $P$ in the tunneling regime.

\section{Role of atomic disorder\label{disorder_geometries}}

In the previous section we have seen that the $3d$ orbitals play an important role
in transport. These orbitals are rather localized on the atoms and the energy 
bands that they give rise to have relatively flat dispersion relations. Therefore,
one would expect the contribution of these orbitals to the transport to be 
very sensitive to the contact geometry. Indeed, in the previous section we have
seen examples in which, by changing the structure of the central part of the
contacts, one can even change the sign of the current polarization. Motivated
by these results, we study in this section in a more systematic manner how
disorder in the atomic positions influences the conductance of one-atom contacts.

In order to simulate the role of disorder we have studied the conductance of contacts
in which the atomic positions in the central cluster have been changed randomly using the geometries of 
Figs.~\ref{single-atom} and \ref{dimer} as starting points. In Fig.~\ref{stat} we present
an example of such a study, where we show histograms of the channel-resolved transmissions at the 
Fermi energy $T_{n,\sigma}(E_F)$ for both spins $\sigma$ constructed from around 3000 
realizations of disorder for each contact. 
The amplitude of the random displacement in each direction was in this case 0.05 times 
the lattice constant. Similar results for contacts of the noble metal Au are also 
shown for comparison. Moreover, for the ferromagnetic materials the insets show 
corresponding histograms of the spin polarization $P$ of the current. 

Let us now discuss the main features of the transmission histograms. First, they show 
that the number of channels obtained for the ideal geometries in the previous 
section is robust with respect to disorder, 
although the transmission coefficients depend crucially on the precise atomic positions. Second,
for the minority spins one has a non-negligible contribution of at least 5 
channels, which originate mainly from the $d$ bands. For the majority spins the
number of channels is clearly smaller and is progressively reduced as we go from Fe to Ni.
This is particularly obvious in the panels of the dimer structures, where one can see that for 
Fe there are three sizable channels and the contribution of the smallest two decreases
for Co and Ni. As explained in the previous section, this is a consequence of the
relative position of the Fermi energy in these three metals. For the latter case of Ni, 
one channel clearly dominates the majority spin conductance, but second 
and third channels are still present. 
Thus, unlike in the case of noble metals such as Au, which only have a single channel 
(see Fig.~\ref{stat}), for ferromagnetic materials conductance quantization is not expected.
Third, the peaks in the histograms for the 
ferromagnetic metals are much broader (especially for the minority spins) than for Au. 
This is due to the higher sensitivity of the $d$ bands to the atomic positions, as compared to 
the $s$ orbitals that dominate the transport in the case of Au. This higher sensitivity is 
a result of the anisotropic spatial dependence of the $d$ orbitals.

In addition, we have calculated the values of the current 
polarization $P$ for each realization of disorder in the contacts. The resulting 
histograms can be found as insets in the panels of Fig.~\ref{stat}. The peaks 
in each histogram are centered around the polarization values of the corresponding ideal 
geometries of Sec.~\ref{ideal_geometries}.

To end this section we would like to make the following comment. In this work
we have analyzed the conductance of some ideal one-atom geometries and the
influence of disorder. These types of calculations are very
valuable to elucidate the nature of the electrical conduction in atomic wires.
However, one has to be cautious in establishing a direct comparison between such theoretical
results and the experiments because of the lack of knowledge of the exact geometries 
realized experimentally. Ideally, the theory should aim at describing the conductance
histograms, which contain the full experimental information without any selection of
the data. This is precisely what we have done for Ni contacts in our recent work 
[\onlinecite{Pauly2006}] and refer the reader to it for further details.

\section{Conclusions\label{conclusions}}

In this work we have presented a theoretical analysis of the conductance of
one-atom thick contacts of the ferromagnetic 3$d$ metals Fe, Co, and Ni. Our calculations are based
on a self-consistent tight-binding model that has previously been successful in
describing the electrical conduction in non-magnetic atomic-sized contacts. Our results indicate
that the $d$ orbitals of these transition metals play a fundamental role in the transport,
especially for the minority-spin species. In the case of one-atom contacts, these orbitals 
combine to provide several partially open conduction channels, which has the following 
important consequences. First, there is no conductance quantization, neither integer nor
half-integer. Second, the current in these junctions is, in general, not fully 
spin-polarized. Third, the conductance of the last plateau is typically above $G_0$.
Finally, both the conductance and the spin polarization of the current are very 
sensitive to the contact geometry. The ensemble of these theoretical findings supports
the recent observations of Untiedt \emph{et al.},\cite{Untiedt2004} while it is in 
clear contradiction with the observations of half-integer conductance 
quantization.\cite{Elhoussine2002,Shimizu2002,Gillingham2002,Rodrigues2003,Gillingham2003a,Gillingham2003b}

It is interesting to mention that in the tunnel regime, when the contacts are actually
broken, the nature of the conduction changes radically. We have shown that in this
case the transport is mainly dominated by the $s$ orbitals and the spin polarization
of the current can reach values close to +100\%.

We want to stress that in all our calculations we have assumed that the atomic contacts 
were formed by single magnetic domains. In this sense, it would be interesting to see
how the conductance in these calculations is modified by the presence of domain walls
in the contact region. The first theoretical studies~\cite{Smogunov2002,Velev2004,
Bagrets2004,Jacob2005} along these lines show that the presence of a domain wall cannot 
conclusively explain the appearance of huge magnetoresistance values reported in the 
literature.~\cite{Garcia1999}

Recently, the so-called anisotropic magnetoresistance has been observed
in ferromagnetic atomic contacts.\cite{Viret2002,Keane2006,Bolotin2006b,Viret2006} 
This effect, in other words the dependence of the resistance on the relative alignment 
of the current and the magnetization, stems from the spin-orbit coupling, 
and can give rise to a correction to the resistance on the order of 1\% in bulk 
ferromagnets.\cite{Mcguire1975} Although the correction can be bigger for atomic-sized 
contacts,\cite{Bolotin2006b,Viret2006} it is nevertheless expected to be a relatively small effect. 
The main ingredient that determines the conduction in the $3d$ ferromagnets is the 
electronic structure, which is what we have described in this work.

\section{Acknowledgments}

We thank Elke Scheer, Magdalena H\"ufner, S\"oren Wohlthat, and Michel Viret for helpful 
discussions. MH, JKV, FP, and JCC were supported financially by the Landesstiftung 
Baden-W\"{u}rttemberg within the {}``Kompetenznetz Funktionelle Nanostrukturen'', 
the Helmholtz Gemeinschaft within the {}``Nachwuchsgruppen-Programm'' (Contract No.~VH-NG-029) 
and the DFG within the CFN. DF was supported by the European Commission through the
Research Training Network (RTN) ``Spintronics". MD and PN appreciate the support by the 
SFB~513.



\begin{thebibliography}{}
\bibitem{Agrait2003}
N. Agra\"{\i}t, A. Levy Yeyati, and J.M. van Ruitenbeek, Phys. Rep. {\bf  377}, 81 (2003).

\bibitem{Cuevas1998a}
J.C. Cuevas, A. Levy Yeyati, and A. Mart\'{\i}n-Rodero, Phys. Rev. Lett. {\bf  80}, 1066 (1998).

\bibitem{Scheer1998}
E. Scheer, N. Agra\"{\i}t, J.C. Cuevas, A. Levy Yeyati, B. Ludoph, A. Mart\'{\i}n-Rodero,
G. Rubio, J.M. van Ruitenbeek and C. Urbina, Nature {\bf  394}, 154 (1998).

\bibitem{Cuevas1998b}
J.C. Cuevas, A. Levy Yeyati, A. Mart\'{\i}n-Rodero, G. Rubio Bollinger,
C. Untiedt, and N. Agra\"{\i}t, Phys. Rev. Lett. {\bf  81}, 2990 (1998).

\bibitem{Ludoph2000a}
B. Ludoph, N. van~der Post, E.N. Bratus', E.V. Bezuglyi, V.S. Shumeiko,
G. Wendin, and J.M. van Ruitenbeek, Phys. Rev. B {\bf  61}, 8561 (2000).

\bibitem{Sirvent1996}
C. Sirvent, J.G. Rodrigo, S. Vieira, L. Jurczyszyn, N. Mingo, and F. Flores,
Phys. Rev. B {\bf  53}, 16086 (1996).

\bibitem{Costa1997}
J.L. Costa-Kr{\"a}mer, Phys. Rev. B {\bf  55}, R4875, (1997).

\bibitem{Hansen1997}
K. Hansen, E. Laegsgaard, I. Stensgaard, and F. Besenbacher,
Phys. Rev. B {\bf  56}, 2208 (1997).

\bibitem{Ott1998}
F. Ott, S. Barberan, J.G. Lunney, J.M.D. Coey, P. Berthet, A.M.
de Leon-Guevara, and A. Revcolevschi, Phys. Rev. B {\bf  58}, 4656 (1998).

\bibitem{Oshima1998}
 H. Oshima and K. Miyano, Appl. Phys. Lett. {\bf  73}, 2203 (1998).

\bibitem{Ono1999}
T. Ono, Y. Ooka, H. Miyajima, and Y. Otani, Appl. Phys. Lett. {\bf  75}, 1622 (1999).

\bibitem{Komori1999}
F. Komori and K. Nakatsuji, J. Phys. Soc. Jap. {\bf  68}, 3786 (1999).

\bibitem{Garcia1999}
N. Garc\'{\i}a, M. Mu{\~n}oz, and Y.-W. Zhao, Phys. Rev. Lett. {\bf  82}, 2923 (1999).

\bibitem{Ludoph2000b}
B. Ludoph and J.M. van Ruitenbeek, Phys. Rev. B {\bf  61}, 2273, (2000).

\bibitem{Viret2002}
M. Viret, S. Berger, M. Gabureac, F. Ott, D. Olligs, I. Petej, J.F. Gregg, C. Fermon, 
G. Francinet, and G. Le Goff, Phys. Rev. B {\bf  66}, 220401(R) (2002).

\bibitem{Elhoussine2002}
F. Elhoussine, S. M{\'a}t{\'e}fi-Tempfli, A. Encinas, and L. Piraux, 
Appl. Phys. Lett. {\bf  81}, 1681 (2002).

\bibitem{Shimizu2002}
M. Shimizu, E. Saitoh, H. Miyajima, and Y. Otani, J. Magn. Magn. Mat. {\bf  239},
243 (2002).

\bibitem{Gillingham2002}
D. Gillingham, I. Linington, and J. Bland, J. Phys.: Condens. Matter {\bf  14}, L567 (2002).

\bibitem{Rodrigues2003}
V. Rodrigues,  J. Bettini, P.C. Silva, and D. Ugarte, 
Phys. Rev. Lett. {\bf  91}, 96801 (2003).

\bibitem{Gillingham2003a}
D. Gillingham, C. M\"uller, and J. Bland, J. Phys.: Condens. Matter {\bf  15}, L291 (2003).

\bibitem{Gillingham2003b}
D. Gillingham, I. Linington, C. M\"uller, and J. Bland, J. Appl. Phys. {\bf  93}, 7388 (2003).

\bibitem{Untiedt2004}
C. Untiedt, D.M.T. Dekker, D. Djukic, and J.M. van Ruitenbeek, Phys. Rev. B {\bf  69},
081401(R) (2004).

\bibitem{Gabureac2004}
M. Gabureac, M. Viret, F. Ott, and  C. Fermon, Phys. Rev. B {\bf  69}, 100401(R) (2004).

\bibitem{Yang2004}
C.-S. Yang, C. Zhang, J. Redepenning, and B. Doudin, Appl. Phys. Lett. {\bf  84},
2865 (2004).

\bibitem{Costa2005}
J.L. Costa-Kr{\"a}mer, M. D\'{\i}az, P.A. Serena, Appl. Phys. A {\bf 81}, 1539 (2005).

\bibitem{Bolotin2006a}
K.I. Bolotin, F. Kuemmeth, A.N. Pasupathy, and D.C. Ralph, Nano Lett. {\bf  6}, 123 (2006).

\bibitem{Keane2006}
Z.K. Keane, L.H. Yu, and D. Natelson, Appl. Phys. Lett. {\bf  88}, 062514 (2006).

\bibitem{Bolotin2006b}
K.I. Bolotin, F. Kuemmeth, and D.C. Ralph, cond-mat/0602251.

\bibitem{Viret2006}
M. Viret, M. Gabureac, F. Ott, C. Fermon, C. Barreteau, G. Autes, and R. Guirardo-Lopez,
Eur. Phys. J. B {\bf 51}, 1 (2006).

\bibitem{Martin2001}
A. Mart\'{\i}n-Rodero, A. Levy Yeyati, and J.C. Cuevas, Physica C {\bf  352}, 67 (2001). 

\bibitem{Smogunov2002}
A. Smogunov, A. Dal Corso, and E. Tossati, Surf. Sci. {\bf  507}, 609 (2002);
{\bf  532}, 549 (2003).

\bibitem{Delin2003}
A. Delin and E. Tosatti, Phys. Rev. B {\bf  68}, 144434 (2003).

\bibitem{Velev2004}
J. Velev and W.H. Butler, Phys. Rev. B {\bf  69}, 094425 (2004).

\bibitem{Bagrets2004}
A. Bagrets, N. Papanikolaou, and I. Mertig, Phys. Rev. B {\bf  70}, 064410 (2004).

\bibitem{Rocha2004}
A.R. Rocha and S. Sanvito, Phys. Rev. B {\bf  70}, 094406 (2004).

\bibitem{Jacob2005}
D. Jacob, J. Fern\'andez-Rossier, and J.J. Palacios, Phys. Rev. B {\bf  71},
220403(R) (2005).

\bibitem{Wierzbowska2005}
M. Wierzbowska, A. Delin, and E. Tosatti, Phys. Rev. B {\bf  72}, 035439 (2005).

\bibitem{Dalgleish2005}
H. Dalgleish and G. Kirczenow, Phys. Rev. B {\bf  72}, 155429 (2005).

\bibitem{Jacob2006}
D. Jacob and J.J. Palacios, Phys. Rev. B {\bf  73}, 075429 (2006).

\bibitem{Mehl1996}
M.J. Mehl and D.A. Papaconstantopoulos, Phys. Rev. B {\bf  54}, 4519 (1996).

\bibitem{Haefner2004}
M. H\"afner, P. Konrad, F. Pauly, J.C. Cuevas, and E. Scheer, 
Phys. Rev. B {\bf  70}, 241404(R) (2004).


\bibitem{Brandbyge1999}
M. Brandbyge, N. Kobayashi, and M. Tsukada, Phys. Rev. B {\bf  60}, 17064 (1999).

\bibitem{Dreher2005}
M. Dreher, F. Pauly, J. Heurich, J.C. Cuevas, E. Scheer, and P. Nielaba, 
Phys. Rev. B {\bf  72}, 075435 (2005).

\bibitem{Viljas2005}
J.K. Viljas, J.C. Cuevas, F. Pauly, and M. H\"afner, Phys. Rev. B {\bf  72},
245415 (2005).

\bibitem{Bacalis2001}
N.C. Bacalis, D.A. Papaconstantopoulos, M.J. Mehl, and M. Lach-hab, Physica B 
{\bf  296}, 125 (2001).

\bibitem{Guinea1983}
F. Guinea, C. Tejedor, F. Flores, and E. Louis, Phys. Rev. B {\bf  28}, 
4397 (1983).

\bibitem{CRC1999}  D.R. Lide, {\it CRC Handbook of Chemistry and Physics, 79th 
Edition} (CRC Press, Boca Raton, Florida, 1998).  

\bibitem{Messiah}  A. Messiah, {\it Quantum Mechanics, Vol. I} 
(Wiley \& Sons, 1958).  

\bibitem{Jelinek2003}
P. Jel\'{\i}nek, R. P\'erez, J. Ortega, and F. Flores, Phys. Rev. B {\bf  68}, 
085403 (2003).

\bibitem{Pauly2006}
F. Pauly, M. Dreher, J.K. Viljas, M. H\"afner, J.C. Cuevas, and P. Nielaba, 
cond-mat/0607129.

\bibitem{Soulen1998}
R.J. Soulen, Jr., J.M. Byers, M.S. Osofsky, B. Nadgorny, T. Ambrose, S.F. Cheng, 
P.R. Broussard, C.T. Tanaka, J. Nowak, J.S. Moodera, A. Barry, and J.M.D. Coey, 
Science {\bf  282}, 85 (1998).

\bibitem{Mcguire1975}
T.R. Mcguire and R.I. Potter, IEEE Trans. Magn. {\bf  11}, 1018 (1975).
 
\end{thebibliography}
\end{document}